\documentclass[12pt]{amsart}
\input{amssymb.sty}
\theoremstyle{plain}

\newtheorem*{thm*}{Theorem}

\newtheorem*{conj*}{Conjecture}

\theoremstyle{definition}
\newtheorem*{defn*}{Definition}

\newtheorem*{rems*}{Remarks}
\newtheorem*{proof*}{Proof}

\newcommand{\nc}{\newcommand}
\nc{\nt}{\newtheorem}
\nc{\gf}[2]{\genfrac{}{}{0pt}{}{#1}{#2}}
\nc{\mb}[1]{{\mbox{$ #1 $}}}
\nc{\real}{{\mathbb R}}
\nc{\comp}{{\mathbb C}}
\nc{\ints}{{\mathbb Z}}
\nc{\Ltoo}{\mb{L^2({\mathbf H})}}
\nc{\rtoo}{\mb{{\mathbf R}^2}}
\nc{\slr}{{\mathbf {SL}}(2,\real)}
\nc{\slz}{{\mathbf {SL}}(2,\ints)}
\nc{\su}{{\mathbf {SU}}(1,1)}
\nc{\so}{{\mathbf {SO}}}
\nc{\hyp}{{\mathbb H}}
\nc{\disc}{{\mathbf D}}
\nc{\torus}{{\mathbb T}}

\newcommand{\CC}{{\mathcal C}}

\nc{\ca}{{\mathcal A}}
\nc{\cag}{{{\mathcal A}^\Gamma}}
\nc{\cg}{{\mathcal G}}
\nc{\chh}{{\mathcal H}}
\nc{\ck}{{\mathcal B}}
\nc{\cl}{{\mathcal L}}
\nc{\cm}{{\mathcal M}}
\nc{\cs}{{\mathcal S}}
\nc{\cz}{{\mathcal Z}}
\nc{\G}{{\mathcal G}}
\nc{\Sc}{{\mathcal S}}
\nc{\E}{{\mathcal E}}

\def\cH{{\mathcal H}}
\def\cK{{\mathcal K}}
\def\cA{{\mathcal A}}
\def\cU{{\mathcal U}}

\def\ZZ{{\mathbb Z}}
\def\RR{{\mathbb R}}
\def\CC{{\mathbb C}}

\nc{\sind}{\sigma{\rm -ind}}
\begin{document}


\title[$D$-Branes, $B$-fields and twisted $K$-theory]
{$D$-Branes, $B$-fields and twisted $K$-theory}
\author{P. Bouwknegt}
\address{Department of Physics and Mathematical Physics,
University of Adelaide, Adelaide 5005, Australia}
\email{pbouwkne@physics.adelaide.edu.au}
\author{V. Mathai}
\address{Department of Pure Mathematics, University of Adelaide, Adelaide
5005, Australia}
\email{vmathai@maths.adelaide.edu.au}


\begin{abstract}
In this note we propose that $D$-brane charges, in the presence of
a topologically non-trivial $B$-field, are classified by
the $K$-theory of an infinite dimensional $C^*$-algebra.  In the case of
$B$-fields whose curvature is pure torsion our description
is shown to coincide with that of Witten.
\end{abstract}

\maketitle

\section{Introduction and discussion}

Recently it was realized that, as a consequence of the fact that
$D$-branes naturally come equipped with (Chan-Paton) vector bundles,
$D$-brane charges take values in the $K$-theory of the spacetime
manifold $X$, rather than in the integral cohomology
$H^\bullet(X,\ZZ)$, as one naively might have expected \cite{MM,Wi}.
In fact, it has been argued that the RR-fields themselves are classified
by the $K$-theory of $X$ as well \cite{MW}.

The identification of $D$-brane charges and $K$-theory classes
has, among other things, led to a better understanding of the spectrum
of $D$-branes, in particular of the existence of stable, nonsupersymmetric
(i.e., non-BPS) $D$-branes \cite{Se}.  In many cases these novel
stable, nonsupersymmetric $D$-branes can be understood as bound states
of a brane-antibrane system with tachyon condensation.
In particular, brane-antibrane
annihilation in the case of $D9$-branes \cite{Se} has been an
important tool (and motivation) in Witten's work \cite{Wi}.

The identification of $D$-brane charges and $K$-theory discussed
above holds in the case of a vanishing NS $B$-field.
In the presence of a $B$-field the arguments of \cite{Wi} need to be
modified as is apparent, for instance, from the analysis of global
string worldsheet anomalies \cite{Wi,FW,Ka}.
In addition, it is well-known that gauge fields on the
$D$-brane in the presence of a (constant) $B$-field are more naturally
interpreted as connections over noncommutative algebras rather than as
connections on vector bundles (see, e.g., \cite{CDS,SW}
and references therein).  Therefore it is natural to suspect
that $D$-brane charges in the presence of a $B$-field should be
identified with the $K$-theory of some noncommutative algebra.  Recall that
$B$-fields are topologically classified by the cohomology class of their
field strength $H$, i.e., $[H]\in H^3(X,\ZZ)$
(see, e.g., Section 6 in \cite{FW} for a mathematical treatment
of $B$-fields).  In the case that $[H]\in Tors(H^3(X,\ZZ))$, i.e.\ $[H]$
represents a torsion class in $H^3(X,\ZZ)$ (the case
of a flat $B$-field), Witten has argued that
the $D$-brane charges take values in a certain twisted version of
$K$-theory (see \cite{Wi}, Section 5.3) or, equivalently, in the $K$-theory
of a certain noncommutative algebra over $X$, known as an Azumaya algebra.
This proposal has been worked out and analyzed in more detail in \cite{Ka}.
What happens in the case when $[H]\neq0$ is not torsion, i.e.\ in the
case of $D$-branes in the presence of NS-charged backgrounds, has remained
obscure so far.

The purpose of this note is to propose a candidate for the relevant
$K$-theory in the more general case.  We will argue that $D$-brane
charges, in the presence of a topologically non-trivial $B$-field, are
classified by the twisted $K$-theory of certain infinite-dimensional,
locally trivial, algebra bundles of compact operators as introduced by
Dixmier and Douady \cite{DD}.  The necessity of going to infinite
dimensional algebra bundles, in order to incorporate nontorsion
classes can, in a sense, be interpreted as going off-shell (as
anticipated in Section 5.3 of \cite{Wi}).  The relevant twisted
$K$-theory, which was defined by Rosenberg \cite{Ros}, is however
not much `bigger' than its finite-dimensional counterpart.  In fact,
we will show that even though the underlying $C^*$-algebra is infinite
dimensional, its twisted $K$-theory has a finite dimensional (local
coordinate) description.  Furthermore, we will show that for $[H] \in
Tors(H^3(X,\ZZ))$ our proposal is equivalent to that of \cite{Wi,Ka}.
The potential relevance of Dixmier-Douady theory to the classification
of $D$-brane charges in the presence of topologically non-trivial
$B$-fields was already noticed in \cite{Ka}.  However, in that paper
this possibility was dismissed for the wrong reasons.

In the remainder of this section we briefly summarize our proposal.
The relevant mathematical constructions and background material
will be explained in more detail
in the next few sections.  A more detailed exposition of our proposal
and its relevance for string theory will appear elsewhere
\cite{BM}.

The starting point of the Sen-Witten construction of stable
nonsupersymmetric $D$-branes (in type IIB String Theory), as remarked above,
is a configuration of $n$ $D9$ brane-antibrane pairs.
When $[H]=0$ the $D$-branes carry a principal $U(n)$ bundle, while
for $[H]\neq0$ the $D$-branes carry a principal $SU(n)/\ZZ_n
= U(n)/U(1)$ bundle over $X$.
Cancelation of global string worldsheet anomalies, however, requires $n[H]=0$
\cite{Wi,Ka,FW}, i.e., requires $[H]$ to be a torsion element.
To incorporate nontorsion $[H]$ we somehow need to take the limit $n\to
\infty$.  This leads us to the study of principal $PU(\cH)=
U(\cH)/U(1)$ bundles over $X$ where $\cH$ is an infinite dimensional,
separable, Hilbert space.  [In fact, $\lim_{n\to\infty} SU(n)/\ZZ_n
= PU(\cH)$ in a certain sense which will be made more precise in the paper.]
It turns out that isomorphism classes of principal $PU(\cH)$ bundles over
$X$ are parametrized precisely by $H^3(X,\ZZ)$.  If we denote by $\cK$
the $C^*$-algebra of compact operators on $\cH$ one can identify
$PU(\cH) = Aut(\cK)$.  It follows that isomorphism classes of locally
trivial bundles over $X$ with fibre $\cK$ and structure group $Aut(\cK)$
are also parametrized by $H^3(X,\ZZ)$.  The cohomology class that is
associated to a locally trivial bundle $\mathcal E$ over $X$ with
fibre $\cK$ and structure group $Aut(\cK)$ is called the Dixmier-Douady
invariant of $\mathcal E$ and is denoted by $\delta(\mathcal E)$.

Our proposal is that $D$-brane charges, in the presence of nontrivial
$[H] \in H^3(X,\ZZ)$, are classified by the $K$-theory of the
$C^*$-algebra $C_0(X,{\mathcal E}_{[H]})$ of continuous
sections that vanish at infinity, of the algebra bundle ${\mathcal E}_{[H]}$
over $X$ with fibre $\cK$ and structure group $Aut(\cK)$ where
$\delta({\mathcal E}_{[H]})=[H]$, i.e., by
\begin{equation}
K^j(X,[H]) ~=~ K_j(C_0(X,{\mathcal E}_{[H]})) \,,\qquad j=0,1\,.
\end{equation}
This is precisely the twisted $K$-theory as defined by Rosenberg \cite{Ros}.

In the case when $[H]$ is a torsion element in $H^3(X,\ZZ)$ the
algebra $C_0(X,{\mathcal E}_{[H]})$, defined above, is Morita
equivalent to the Azumaya algebra defined in \cite{Wi,Ka} and
therefore the $K$-theories of these algebras are canonically
isomorphic.

Note that our proposal answers some of the questions raised at the end
of Section 5.3 in \cite{Wi} and refutes some of the remarks made in
Section 7 of \cite{Ka}, concerning the case when the NS 3-form field
$H$ is a not a torsion element in $H^3(X,\ZZ)$. We also mention that
twisted $K$-theory has appeared earlier in the mathematical physics
literature, in the study of the quantum Hall effect \cite{CHMM,CHM}.


\section{Brauer groups and the Dixmier-Douady invariant}

Let $X$ be a locally compact, Hausdorff space with a countable basis of
open sets, for example a smooth manifold. Then recall that the classifying
space of the third cohomology group of $X$
is the Eilenberg-Maclane space $K(\ZZ,3)$,
where $K(\ZZ,k)$ is defined uniquely upto homotopy as being the
topological space  with the property that $\pi_k(K(\ZZ,k), *) = \ZZ$ and
$\pi_j(K(\ZZ,k), *) = 0$ for $j\ne k$  (cf.\ \cite{Wh}). We will now
describe a candidate for the Eilenberg-Maclane space $K(\ZZ,3)$.

Let $\cH$ denote an infinite dimensional,
separable, Hilbert space and $\cK$ the $C^*$-algebra of compact operators on
$\cH$.  Let $U(\cH)$ denote the group of unitary operators on $\cH$. Then
it is a fundamental theorem of Kuiper that $U(\cH)$ is contractible in the
strong operator topology. Now define the projective unitary group on
$\cH$ as $PU(\cH)= U(\cH)/U(1)$, where $U(1)$ consists of scalar multiples
of the identity operator on $\cH$ of norm equal to 1. It follows that
a model for the classifying space of $U(1)$ is $BU(1) = PU(\cH)$. Since
$\mathbb R$ is contractible, it follows that $U(1) =  \mathbb R/\ZZ$ is
itself a $K(\ZZ, 1)$. Therefore $PU(\cH)$ is a model for $K(\ZZ,2)$
and finally a model for $K(\ZZ, 3)$ is the classifying space of
$PU(\cH)$, i.e., $K(\ZZ,3) = BPU(\cH)$. We conclude that
\begin{equation}
H^3(X, \ZZ) = [X, K(\ZZ,3)] = [X, BPU(\cH)],
\end{equation}
where $[X, Y]$ denotes the homotopy classes of maps from $X$ to $Y$.
In other words, isomorphism classes of principal $PU(\cH)$ bundles
over $X$ are parametrized by  $H^3(X, \ZZ) $. Now for $g\in U(\cH)$,
let $Ad(g)$ denote the automorphism of $\cK$ given by $T\to gTg^{-1}$.
It is well known that $Ad$ is a continuous homomorphism of $U(\cH)$ onto
$Aut(\cK)$ with kernel $U(1)$, where $Aut(\cK)$ is given the point-norm
topology (cf.\ \cite{RW}, Chapter~1).
It then follows that $PU(\cH) = Aut(\cK)$.
Therefore the isomorphism
classes of locally trivial bundles over $X$ with fibre
$\cK$ and structure group
$Aut(\cK)$ are also parametrized by  $H^3(X, \ZZ) $. Since $\cK\otimes
\cK \cong \cK$, we see that the isomorphism
classes of locally trivial bundles over $X$ with fibre $\cK$ and
structure group
$Aut(\cK)$ form a group under the tensor product, where the inverse
of such a bundle is the conjugate bundle. This group is known as the
{\em infinite Brauer group} and is denoted by $Br^\infty(X)$ (cf.\
\cite{Pa}). The cohomology
class in $H^3(X, \ZZ)$ that is associated to a locally trivial
bundle $\mathcal E$ over $X$ with fibre $\cK$ and structure group
$Aut(\cK)$ is called the
{\em Dixmier-Douady invariant} of $\mathcal E$ and is denoted by
$\delta(\mathcal E)$, see \cite{DD}.

We next give a ``local coordinate'' description of elements in
$Br^\infty(X)$.  Let $\{ \cU_i\}$ be a good open cover of $X$, i.e.\
such that all $\cU_i$ and their multiple overlaps are contractible. An
element of the $C^*$-algebra $C_0(X, \mathcal E)$, of continuous sections
of ${\mathcal E}$ that vanish at infinity, is a collection of
functions $R_i: \cU_i\to \cK$ such that on the overlaps $\cU_i\cap
\cU_j$ one has
\begin{equation}\label{a}
R_i=g_{ij} R_j g_{ij}^{-1} = Ad(g_{ij}) R_j.
\end{equation}
Here $g_{ij} : \cU_i\cap \cU_j \to U(\cH)$ are continuous functions on
the overlaps, satisfying $g_{ij} g_{ji} =1$, and therefore $Ad(g_{ij}) :
\cU_i\cap
\cU_j
\to PU(\cH) = Aut(\cK)$.
Consistency of (\ref{a}) on triple overlaps $\cU_i\cap \cU_j\cap
\cU_k$ implies that
\begin{equation}\label{b}
g_{ij} g_{jk} g_{ki}=\zeta_{ijk},
\end{equation}
where $\zeta_{ijk}$ are $U(1)$-valued functions. One verifies
that on quadruple overlaps $\cU_i\cap
\cU_j\cap \cU_k \cap \cU_l$, the functions
$\zeta_{ijk}$ satisfy
\begin{equation}
\zeta_{ijk}\zeta_{ikl} = \zeta_{jkl}\zeta_{ijl}.
\end{equation}
Therefore we see that on quadruple overlaps $\cU_i\cap
\cU_j\cap \cU_k \cap \cU_l$, one has
\begin{equation}
\log \zeta_{ijk} + \log \zeta_{ikl} - \log \zeta_{jkl}
- \log \zeta_{ijl} = 2\pi \sqrt{-1} \kappa_{ijkl}
\end{equation}
where $\{\kappa_{ijkl}\}$ is a $\ZZ$-valued \v Cech 3-cocycle, and
therefore defines an element $\kappa \in H^3(X,\ZZ)$. This
is the Dixmier-Douady class $\;\delta(\mathcal E)$ mentioned in the
paragraph above.

Let $Tors(H^3(X, \ZZ) )$ denote the subgroup of torsion elements in
$H^3(X, \ZZ) $. Suppose now that $X$ is compact.  Then there is a
well known description of $Tors(H^3(X, \ZZ) )$ in terms of finite dimensional
Azumaya algebras over $X$ \cite{DK}. Recall that
an Azumaya algebra of rank $m$ over $X$ is a locally trivial algebra bundle
over $X$ whose fibre is isomorphic to the algebra of $m\times m$ matrices
$M_m(\CC)$.
An example of an Azumaya algebra over $X$ is
 the algebra ${\rm End}(E)$ of all endomorphisms of a vector bundle $E$ over
$X$. Two Azumaya algebras $\mathcal E$ and $\mathcal F$
over $X$ are said to be
equivalent if there are vector bundles $E$ and $F$ over $X$ such that
$\mathcal E\otimes {\rm End}(E)$ is isomorphic to
$\mathcal F\otimes {\rm End}(F)$. In particular, an Azumaya algebra of the
form
${\rm End}(E)$ is equivalent to $C(X)$ for any
vector bundle $E$ over $X$.
The group of all equivalence classes of Azumaya algebras over $X$ is
called the
Brauer group of $X$ and is denoted by $Br(X)$. We will denote by
$\delta'(\mathcal E)$ the class in $Tors(H^3(X,\ZZ))$
corresponding to the Azumaya algebra $\mathcal E$ over $X$.
It is constructed by analogy to the local coordinate
description given in the previous paragraph.
Serre's theorem asserts that $Br(X)$ and $Tors(H^3(X,\ZZ))$
are isomorphic.

Thus we see that there are two distinct descriptions of $Tors(H^3(X, \ZZ) )$,
one in terms of finite dimensional Azumaya algebras over $X$, and
the other is terms of locally trivial bundles over $X$ with fibre
$\cK$ and structure group $Aut(\cK)$. These two descriptions are related as
follows. Given an Azumaya algebra $\mathcal E$ over $X$, then the tensor
product $\mathcal E\otimes \cK$ is a locally trivial bundle over $X$ with
fibre $M_m(\CC)\otimes~\cK\cong \cK$ and
structure group $Aut(\cK)$, such that $\delta'(\mathcal E)
= \delta(\mathcal E\otimes \cK)$. Notice that the algebras
$C(X, \mathcal E)$ and
$C(X, \mathcal E\otimes \cK) = C(X, \mathcal E)\otimes \cK$ are
Morita equivalent. Moreover if
$\mathcal E$ and  $\mathcal F$ are equivalent  Azumaya algebras
over $X$, then $\mathcal E\otimes \cK$  and $\mathcal F\otimes
\cK$ are isomorphic locally trivial bundles over $X$ with fibre
$\cK$ and structure group $Aut(\cK)$. To see this, we recall
that $\cK = \lim_{n} M_n(\CC)$ where the limit is taken in the
$C^*$-norm topology \cite{Dix}. Since the automorphism group of
$M_n(\CC)$
is $PU(n) = SU(n)/\ZZ_n$ and $Aut(\cK) = PU(\cH)$, it is in this
sense that $\lim_{n\to\infty} SU(n)/\ZZ_n = PU(\cH)$.
The equivalence relation
for the Azumaya algebras $\mathcal E$ and  $\mathcal F$ becomes,
$\mathcal E\otimes \cK(E)$ and  $\mathcal F \otimes \cK(F)$
are isomorphic, where $\cK(E)$ and $\cK(F)$ are the bundles of
compact operators on the infinite dimensional Hilbert bundles
$E$ and $F$, respectively. By Kuiper's theorem, the group $U(\cH)$
of unitary operators in an infinite dimensional Hilbert space
$\cH$ is contractible in the strong operator
topology. Therefore, the infinite dimensional Hilbert bundles
$E$ and $F$ are trivial, and therefore both $\cK(E)$ and $\cK(F)$
are isomorphic to $X\times \cK$. It follows that $\mathcal
E\otimes \cK(E)$ and  $\mathcal F \otimes \cK(F)$
are isomorphic if and only if $\mathcal E\otimes \cK$ and
$\mathcal F \otimes \cK$ are isomorphic, as asserted.

Now, suppose that
$\mathcal E$ is a locally trivial bundle over $X$ with fibre
$\cK$ and structure group $Aut(\cK)$, such that
$\delta(\mathcal E)$ is a torsion element, then there is a
positive integer $n$ such that $0 = n \delta(\mathcal E)
=  \delta({\mathcal E}^{\otimes n})$. Therefore ${\mathcal
E}^{\otimes n}$ is isomorphic to the trivial bundle $X\times
\cK$, and it follows that $\mathcal E$  has transition
functions $g_{ij}: \cU_i\cap\cU_j \to Aut(\cK)$ that are locally
constant functions. That is, $\mathcal E$ is a {\em flat}
locally trivial bundle over $X$ with fibre
$\cK$ and structure group $Aut(\cK)$, which is given by a
representation of $\pi_1(X)$ into  $Aut(\cK)$. Therefore we
see that  $Tors(H^3(X, \ZZ) )$ parametrizes the topologically
nontrivial isomorphism classes of {\em flat} locally trivial
bundles over $X$ with fibre
$\cK$ and structure group $Aut(\cK)$. In fact, given a representation
$\rho: \pi_1(X) \to Aut(\cK) = PU(\cH)$, there is a map
$\lambda: \pi_1(X) \to U(\cH)$, such that $\rho(\gamma) =
Ad(\lambda(\gamma))$ for all $\gamma \in  \pi_1(X)$,
satisfying the identity
\begin{equation}
\lambda(\gamma_1) \lambda(\gamma_2)
\lambda(\gamma_1 \gamma_2)^{-1} = \sigma(\gamma_1, \gamma_2),
\qquad \forall \gamma_1, \gamma_2 \in  \pi_1(X),
\end{equation}
where $\sigma : \pi_1(X) \times \pi_1(X) \to U(1)$ satisfies
the cocycle identity
\begin{equation} \label{coc}
\sigma(\gamma_1, \gamma_2) \sigma(\gamma_1 \gamma_2, \gamma_3) =
\sigma(\gamma_2, \gamma_3) \sigma(\gamma_1,
\gamma_2 \gamma_3), \qquad \forall \gamma_1, \gamma_2, \gamma_3
\in  \pi_1(X)
\end{equation}
and is normalized, $ 1=\sigma(1, \gamma) =  \sigma(\gamma, 1), \quad
\forall \gamma\in  \pi_1(X)$.
Such a normalized $U(1)$-valued group 2-cocycle on $ \pi_1(X)$
is called a {\em multiplier} on $ \pi_1(X)$.  The flat bundle 
defined by $\rho$ is ${\mathcal E}_\rho = \left(\widetilde X \times
\mathcal K\right)/\sim$, where $(x, T) \sim (\gamma^{-1}x, \rho(\gamma)T)$
for all $\gamma \in \pi_1(X)$ and $\widetilde X$ is the universal cover of
$X$. Then 
${\mathcal E}_\rho$ has Dixmier-Douady invariant $\delta({\mathcal
E}_\rho) =
\delta''(f^*\sigma)$ \cite{Was}, where $\delta''$ is the connecting
homomorphism in the ``change of coefficients'' long exact sequence 
$$
\cdots \to H^2(\pi_1(X), \RR) \to  H^2(\pi_1(X), U(1))
\stackrel{\delta''}{\to} H^3(\pi_1(X),\ZZ) \to  \cdots
$$
that is associated to the short exact sequence of coefficient groups,
$$
0\to \ZZ\to \RR\to U(1)\to 0
$$
and $f:X\to B\pi_1(X)$ denotes the continuous map classifying the
universal cover.  

We will now show that a closed, integral, $\pi_1(X)$-invariant
differential 2-form
$\omega$ on the universal cover $\widetilde X$ of $X$
determines such a multiplier on $ \pi_1(X)$. [In the quantum Hall
effect $\omega$ represents the magnetic field, cf.\ \cite{CHMM,CHM}.]

By geometric pre-quantization, there is an essentially unique Hermitian
line bundle
$\mathcal{L}\to\widetilde{X}$ and a Hermitian connection $\nabla$ whose
curvature is $\nabla^2 = i \omega \; (i=\sqrt{-1})$.
Since $\omega$ is $\pi_1(X)$-invariant, one sees that for $\gamma\in\pi_1(X)$,
$\;\gamma^*\nabla$ is also a
Hermitian vector potential for $\omega$,
i.e., $(\gamma^*\nabla)^2 = i \gamma^* \omega
=i \omega$. Now $\gamma^*\nabla - \nabla = iA_\gamma \in
\Omega^1(\widetilde X, \mathbb R)$. Since $i
\omega= \nabla^2 = (\gamma^*\nabla)^2$,
we see that $0 = \nabla A_\gamma + A_\gamma \nabla = dA_\gamma$, i.e.
$A_\gamma$ is a closed $1$-form on the simply-connected manifold
$\widetilde X$.
Therefore it is exact, i.e.
$A_\gamma = d\phi_\gamma \ (*)$,
where $\varphi_\gamma$ is a real-valued smooth function on $\widetilde X$.
It is easy to see that it also satisfies
\begin{enumerate}
\item[(i)] $\varphi_\gamma(x)+\varphi_{\gamma'}(\gamma^{-1} x)-
\varphi_{\gamma'\gamma}(x)$ is independent of $x\in\widetilde{X}$ for all
$\gamma, \gamma' \in \pi_1(X)$;
\item[(ii)] $\varphi_\gamma(x_0)=0$ for some $x_0\in\widetilde{X}$
and for all $\gamma \in \pi_1(X)$.
\end{enumerate}
Equation (i) follows immediately from $(*)$ and (ii)
is a normalization. For $\gamma\in\pi_1(X)$ define $U_\gamma
f(x)=f(\gamma^{-1}\cdot x)\;$ (translations) and $S_\gamma
f(x)=e^{i\varphi_\gamma(x)}f(x)\;$ (phase) and
$T_\gamma=U_\gamma\circ S_\gamma\;$ (magnetic translations).

Then one computes that
\begin{equation}
T_{\gamma_1} T_{\gamma_2}=\sigma(\gamma_1,\gamma_2)
T_{\gamma_1\gamma_2}\,,\qquad\text{for }\gamma_1,\gamma_2\in\pi_1(X) \,,
\end{equation}
where $\sigma :
\pi_1(X)\times\pi_1(X)\to U(1)$ is defined as
$\sigma(\gamma_1, \gamma_2) = e^{-i\phi_{\gamma_1}(\gamma_2^{-1} \cdot
x_0)},\;\;\forall
\gamma_1, \gamma_2\in\pi_1(X)$. It satisfies the cocycle condition
(\ref{coc})
by the associativity of $T$. Thus $\sigma$ is the multiplier
on $\pi_1(X)$ that is determined
by $\omega$ and a choice of base-point $x_0$. Any other choice of
base-point determines a cohomologous multiplier on $\pi_1(X)$.

Conversely, given a multiplier $\sigma$ on $\pi_1(X)$,
consider the Hilbert space
of square summable functions on $\pi_1(X)$,
\begin{equation}
\ell^2(\pi_1(X))=\left\{f:\pi_1(X)\to \CC:
\;\sum_{\gamma\in\pi_1(X)}|f(\gamma)|^2<\infty \right\}\;.
\end{equation}
The  left $\sigma$-regular
representation on $\ell^2(\pi_1(X))$ is defined
as being, $\quad\forall\;\gamma,\gamma'\in \pi_1(X)$,
\begin{eqnarray}
L^{\sigma} &: \pi_1(X) \longrightarrow U(\ell^2(\pi_1(X)))\nonumber\\
(L_{\gamma}^{\sigma}
f)(\gamma') & =f(\gamma^{-1}\gamma')\sigma(\gamma,\gamma^{-1} \gamma').
\end{eqnarray}
It satisfies
$L_{\gamma}^{\sigma} L_{\gamma'}^{\sigma}  = \sigma(\gamma,\gamma')
L_{\gamma\gamma'}^{\sigma}
$  for all $\gamma, \gamma'\in \pi_1(X)$.
That is, the  left $\sigma$-regular
representation $L^{\sigma}$ on $\ell^2(\pi_1(X))$ is a projective
unitary representation. Therefore $Ad(L^{\sigma}) : \pi_1(X)
\longrightarrow PU(\ell^2(\pi_1(X))) = Aut(\cK)$ is a representation
of  $\pi_1(X)$ into $Aut(\cK)$, and so determines a {flat} locally
trivial bundle ${\mathcal E}_\sigma$ over $X$ with fibre
$\cK$ and structure group $Aut(\cK)$. It follows by a result in
\cite{Was} that the Dixmier-Douady invariant of ${\mathcal E}_\sigma$
is  $\delta({\mathcal E}_\sigma) =
\delta''(f^*\sigma)$, where $f$ and $\delta''$ are the same as above.

\section{Twisted $K$-theory and noncommutative geometry}

Let $X$ be a locally compact, Hausdorff space with a countable basis of
open sets, for example a smooth manifold. Let $[H] \in H^3(X, \ZZ)$. Then
the {\em twisted $K$-theory} was defined by Rosenberg \cite{Ros} as
\begin{equation}
K^j(X, [H]) = K_j(C_0(X, {\mathcal E}_{[H]}))\qquad j=0,1,
\end{equation}
where ${\mathcal E}_{[H]}$ is the unique  locally trivial
bundle over $X$ with fibre
$\cK$ and structure group $Aut(\cK)$ such that
$\delta({\mathcal E}_{[H]}) = [H]$,
and $K_\bullet(C_0(X, {\mathcal E}_{[H]}))$ denotes the
topological $K$-theory of the
$C^*$-algebra of continuous sections of ${\mathcal E}_{[H]}$
that vanish at infinity.  Notice that when $[H] = 0 \in H^3(X,
\ZZ)$, then ${\mathcal E}_{[H]} = X\times \cK,$ therefore
$C_0(X, {\mathcal E}_{[H]}) = C_0(X)\otimes \cK$ and
by Morita invariance of $K$-theory, the twisted $K$-theory of
$X$ coincides with  the standard $K$-theory of $X$ in this case.
The Morita invariance of the $K$-theory of a $C^*$-algebra $\cA$
can be explained as follows. $K_0(\cA)$ can be defined as the Grothendieck
group of Murray-von Neumann equivalence classes of projections
in $\cA\otimes \cK$ (cf.\ \cite{WO}).  Therefore the isomorphism
$\cK \otimes \cK \cong \cK$ induces the isomorphism $K_0(\cA)
\cong K_0(\cA\otimes \cK)$. Also $K_1(\cA)$ can be defined as the path
components of the group
$\{g\in U((\cA \otimes \cK)^+): g-1\in \cA\otimes \cK\}$ in the norm
topology,
where $(\cA \otimes \cK)^+$ denotes the $C^*$ algebra obtained from 
$\cA \otimes \cK$ by adjoining the identity operator.
Therefore we again see that the isomorphism
$\cK \otimes \cK \cong \cK$ induces the  isomorphism $K_1(\cA)
\cong K_1(\cA\otimes \cK)$.

When $X$ is compact and when $[H] \in Tors(H^3(X, \ZZ))$, then
there is an alternate description of the twisted $K$-theory
$K^j(X, [H])$, due to Donovan and Karoubi \cite{DK}, as being the topological
$K$-theory of the algebra of sections of an  Azumaya algebra $\mathcal F$ over
$X$ with $\delta'(\mathcal F) = [H]$. Notice that this is well defined as
the $C^*$-algebra of sections over
any other equivalent Azumaya algebra is actually Morita equivalent to $C(X,
\mathcal F)$, and therefore they have the same $K$-theory. The relation
between the Donovan-Karoubi twisted $K$-theory and the Rosenberg twisted
$K$-theory is obtained by tensoring the Azumaya algebra with $\cK$, as
discussed in Section 2, and by Morita invariance of
$K$-theory we see that the two definitions for the twisted
$K$-theory of $X$ are isomorphic .

There are
alternate descriptions of  $K^0(X, [H])$, whose elements are
generated by projections in
$C_0(X, \mathcal E_{[H]})\otimes \cK \cong C_0(X, \mathcal E_{[H]})$.
It is proved in \cite{Ros} that when $X$ is
compact one has
\begin{equation}
K^0(X, [H]) = [Y, U(\mathcal Q)]^{Aut(\cK)}
\end{equation}
where $Y$ is the principal $Aut(\cK)$ bundle over $X$ such that
$\mathcal E = Y \times_{Aut(\cK)} \cK $ and $U(\mathcal Q)$ denotes
the group of unitary elements in the Calkin algebra $\mathcal Q =
B(\cH)/\cK$ in the norm topology, where $B(\cH)$ denotes the algebra
of all bounded linear operators on $\cH$. Another
result in \cite{Ros} is that
when $X$ is compact one has
\begin{equation}
K^1(X, [H]) ~=~ [Y, U(\cH)]^{Aut(\cK)}\,.
\end{equation}

We will next give a ``local coordinate''
description of objects in the twisted $K$-theory.

Since projections in $\cK$ have finite dimensional ranges \cite{Dix},
we observe that even though the algebra $C_0(X, {\mathcal E}_{[H]})$
is infinite dimensional, the projections in $C_0(X, {\mathcal E}_{[H]})$
have ranges which are bundle like objects with finite dimensional fibres.
These can be described as follows. Assume first that $X$ is compact. Recall
that projection $P$ in the the algebra $C_0(X, {\mathcal E}_{[H]})$
satisfies $P^*= P = P^2$. That is, $P$ is a
is a collection of continuous
functions $P_i: \cU_i\to \cK$ such that $P_i^* = P_i
= P_i^2$ and  such that on the overlaps $\cU_i\cap
\cU_j$, one has
\begin{equation}
P_i = Ad(g_{ij}) P_j.
\end{equation}
Here the continuous functions $Ad(g_{ij}) : \cU_i\cap \cU_j \to PU(\cH) =
Aut(\cK)$ are the transition  functions for the locally trivial bundle
${\mathcal E}_{[H]}$ with fibre
$\cK$, where $g_{ij} : \cU_i\cap \cU_j \to U(\cH)$ are continuous functions on
the overlaps, satisfying $g_{ij} g_{ji} =1$, and equation (\ref{b}). Observe
that $(Ad(g_{ij})P_j)^* = Ad(g_{ij})P_j = (Ad(g_{ij})P_j)^2$.  We see
that  the range of the projection $P_i(x)$ is a {\em finite dimensional}
subspace $V_{i,x} \subset \cH$ for each $x\in \cU_i$, and  the collection
$\{V_{i,x}\}_{x\in \cU_i}$ is continuous
over $\cU_i$ in the sense that
$\cU_i \ni x\to P_i(x)$ is continuous. On the overlaps $ \cU_i\cap
\cU_j
$, an element
$v \in  V_{i,x}$ is identified with the element $g_{ji}(x) v \in
V_{j,x}$. We will then
say that the data $\{\cU_i, \{V_{i,x}\}_{x\in \cU_i}, g_{ij}\}$ defines a {\em
gauge-bundle} over
$X$, to be consistent with terminology in the physics literature.
The definition is
incomplete in the sense that it does not discuss the dependence on the
choices made. However this can be remedied by introducing sheaves of
categories \cite{Br} and will be done elsewhere. Note that a gauge-bundle is
not a manifold in general, unlike the definition of a vector bundle.
Therefore, we see that a projection $P$ in $C_0(X, {\mathcal E}_{[H]})$
defines a gauge-bundle over $X$. Recall that two projections $P$ and $Q$
in $C_0(X, {\mathcal E}_{[H]})$ are Murray-von Neumann equivalent if there is
a $\Lambda \in C_0(X, {\mathcal E}_{[H]})$ such that $P = \Lambda^*\Lambda$
and $Q = \Lambda\Lambda^*$. In local coordinates, this means that there is a
collection  of continuous  functions $\Lambda_i: \cU_i\to \cK$ such that $P_i =
\Lambda_i^* \Lambda_i$ and
$Q_i: \Lambda_i\Lambda_i^*$, such that on the
overlaps $\cU_i\cap
\cU_j$, one has
\begin{equation}
\Lambda_i = Ad(g_{ij}) \Lambda_j.
\end{equation}
In terms of gauge-bundles, this means that if $\{\cU_i, \{V_{i,x}\}_{x\in
\cU_i}, g_{ij}\}$ and $\{\cU_i, \{W_{i,x}\}_{x\in \cU_i}, g_{ij}\}$ are the
gauge-bundles defined by the projections $P$ and $Q$ respectively, then an
{\em isomorphism} of gauge bundles is given by such an element $\Lambda$. Note
that $\Lambda_i (x) : V_{i,x} \to ~W_{i,x}$ is an isomorphism of finite
dimensional vector spaces, with inverse $\Lambda_i^* (x) : W_{i,x} \to
~V_{i,x}$.  The direct sum of gauge-bundles  $\{\cU_i, \{V_{i,x}\}_{x\in
U_i}, g_{ij}\}$ and $\{\cU_i, \{W_{i,x}\}_{x\in \cU_i}, g_{ij}\}$ is again a
gauge-bundle
\begin{equation}
\{\cU_i, \{V_{i,x}\}_{x\in
\cU_i}, g_{ij}\} \oplus \{\cU_i, \{W_{i,x}\}_{x\in \cU_i}, g_{ij}\}
=\left\{\cU_i, \{V_{i,x}\oplus W_{i,x}\}_{x\in \cU_i}, g_{ij}\right\}.
\end{equation}
It corresponds to taking the orthogonal direct sum of the projections
that define the gauge-bundles.
Note that the gauge-bundles $\;\{\cU_i, \{V_{i,x}\}_{x\in
\cU_i}, g_{ij}\}\oplus\{\cU_i, \{W_{i,x}\}_{x\in \cU_i}, g_{ij}\}\;$
and
$ \{\cU_i, \{W_{i,x}\}_{x\in \cU_i}, g_{ij}\}~\oplus
~\{\cU_i, \{V_{i,x}\}_{x\in
\cU_i}, g_{ij}\}\;$ are naturally isomorphic.
Define
${\bf Vect} (X, [H]) $ to be the Abelian semigroup of isomorphism classes
of gauge-bundles $[\{\cU_i, \{V_{i,x}\}_{x\in
\cU_i}, g_{ij}\}]$ with the direct sum operation.
Then the associated  Grothendieck group is just $K^0(X, [H])$.  That is,
if $X$ is compact,
$$
\begin{array}{lcl}
K^0(X, [H])& = & \Big\{[\{\cU_i, \{V_{i,x}\}_{x\in
\cU_i}, g_{ij}\}] - [\{\cU_i, \{W_{i,x}\}_{x\in
\cU_i}, g_{ij}\}]: \{\cU_i, \{V_{i,x}\}_{x\in
\cU_i}, g_{ij}\},\\[+7pt]
& &\{\cU_i, \{W_{i,x}\}_{x\in
\cU_i}, g_{ij}\} \;{\rm are\; gauge\!\!-\!\!bundles\; over} \;X\Big\}
\end{array}
$$
When $X$ is not compact, then $K^0(X, [H])$ consists of isomorphism classes
of triples $(\{\cU_i, \{V_{i,x}\}_{x\in
\cU_i}, g_{ij}\}, \{\cU_i, \{W_{i,x}\}_{x\in
\cU_i}, g_{ij}\}, \Lambda)$, where  $\{\cU_i, \{V_{i,x}\}_{x\in
\cU_i}, g_{ij}\}$ and \\
$\{\cU_i, \{W_{i,x}\}_{x\in
\cU_i}, g_{ij}\}$ are gauge-bundles over $X$ and $\Lambda$ is an isomorphism
between the gauge bundles on the complement of a
compact subset of $X$. There is a more elegant description of such
objects using an analogue of Quillen's formalism \cite{Qu}.
This, and
an analogous description for  $K^1(X,[H])$, will be given elsewhere.

\section{Conclusions}

In this paper we have proposed a natural candidate for the
classification of $D$-brane charges, in the presence of
a topologically non-trivial $B$-field, in terms of
the twisted $K$-theory of certain infinite-dimensional, locally
trivial, algebra bundles of compact operators.  We have also shown
that in the case of torsion elements $[H]$ our proposal is equivalent
to that of Witten \cite{Wi}.
While the necessity of incorpating nontorsion classes
$[H]\in H^3(X,\ZZ)$ in the formalism has forced us to consider
infinite dimensional algebra bundles, this description is, in some respects,
more natural even for torsion elements.  The reasons are, first
of all, that the bundle
$\mathcal E_{[H]}$ is the {\em unique} locally trivial bundle over
$X$ such that $\delta(\mathcal E_{[H]})=[H]$, while the corresponding
Azumaya algebras are only determined upto equivalence.  And, secondly,
our proposal holds for any, locally compact, Hausdorff space with a countable
basis of open sets, while the twisted $K$-theory of an Azumaya algebra
is only defined in the case of compact $X$.

It is clear that our proposal needs further study \cite{BM}.  For instance,
by using a fundamental theorem of Grothendieck \cite{Gr},
it can be shown that in the case of torsion elements $[H]$, the groups
$K^\bullet(X)$ and $K^\bullet(X,[H])$ are rationally equivalent
\cite{Wi}.  The proof appears to break down for nontorsion elements,
however.  The structure of
$K(X,[H])$, as well as its physical interpretation, would be
greatly elucidated by studying the Connes-Chern map $K_\bullet(\mathcal A)
\to HPC_\bullet(\mathcal A)$ (see \cite{Co}) and a possible relation of the
periodic cyclic
homology $HPC_\bullet(\mathcal A)$ to the de-Rham cohomology for the
noncommutative algebra $\mathcal A = C_0^\infty(X,\mathcal E_{[H]})$
of smooth sections vanishing to all orders at infinity. Note that by the
analogue
of Oka's  principle in this context, one has $K_\bullet(\mathcal A) \cong
K^\bullet(X, [H])$.

Other issues, such as the cancellation of global string
worldsheet anomalies, discrete torsion in this setting
and examples of $D$-branes in NS-charged backgrounds remain to be
worked out as well. \bigskip

\noindent {\bf Acknowledgements:} P.B.\ and V.M.\ were financially
supported by the Australian Research Council.  We would like to thank
Anton Kapustin for pointing out an error in the local coordinate 
description of the twisted $K$-theory in an earlier version of the 
paper.



\begin{thebibliography}{CHMM}

\bibitem[BM]{BM} P.~Bouwknegt and V.~Mathai, work in progress.

\bibitem[Br]{Br} J.-L.~Brylinski, {\em Loop spaces, characteristic
classes and geometric quantization},
Progress in Mathematics {\bf 107}, (Birkh\"auser, Boston, 1993).

\bibitem[CHMM]{CHMM} A.~Carey, K.~Hannabuss, V.~Mathai and P.~McCann,
{\em Quantum Hall Effect on the hyperbolic plane}, Commun.\ Math.\
Physics, {\bf 190} no. 3 (1997) 629--673, [{\tt dg-ga/9704006}].

\bibitem[CHM]{CHM} A.~Carey, K.~Hannabuss, V.~Mathai, {\em Quantum
Hall effect on the hyperbolic plane in the presence of disorder},
Lett.\ Math.\ Phys.\ {\bf 47} (1999), no. 3,  215--236,
[{\tt math.DG/9804128}].

\bibitem[Co]{Co} A.~Connes, {\em Noncommutative geometry},
(Academic Press, San Diego, 1994).

\bibitem[CDS]{CDS} A.~Connes, M.~Douglas and A.~Schwarz, {\em Noncommutative
geometry and matrix theory: Compactification on tori},
JHEP {\bf 02} (1998) 003, [{\tt hep-th/9711162}].

\bibitem[Dix]{Dix}  J.~Dixmier, {\em C$^*$-algebras and their
representations}, (North Holland, Amsterdam, 1982).

\bibitem[DD]{DD} J.~Dixmier and A.~Douady, {\em Champs continues d'espaces
hilbertiens at de $C^*$-alg\`ebres}, Bull.\ Soc.\ Math.\ France\ {\bf 91}
(1963) 227--284.

\bibitem[DK]{DK} P.~Donovan and M.~Karoubi, {\em Graded Brauer
groups and $K$-theory with local coefficients}, Inst.\ Hautes
\'Etudes Sci.\ Publ.\ Math., {\bf 38} (1970) 5--25.

\bibitem[FW]{FW} D.~Freed and E.~Witten, {\it Anomalies in string theory
with $D$-branes}, [{\tt hep-th/9907189}].

\bibitem[Gr]{Gr} A.~Grothendieck, {\em Le groupe de Brauer, I, II,
III. 1968 Dix Expos\'es sur la Cohomologie des Sch\'emas}, pp. 46--188,
(North-Holland, Amsterdam; Masson, Paris).

\bibitem[Ka]{Ka} A. Kapustin, {\em $D$-branes in a topologically nontrivial
$B$-field}, [{\tt hep-th/9909089}].

\bibitem[MM]{MM} R.~Minasian and G.~Moore, {\em $K$-theory and Ramond-Ramond
charge}, JHEP {\bf 11} (1997) 002, [{\tt hep-th/9710230}].

\bibitem[MW]{MW} G.~Moore and E.~Witten, {\em Self-duality, Ramond-Ramond
fields, and $K$-theory},\newline [{\tt hep-th/9912279}].

\bibitem[Pa]{Pa} E.M.~Parker, {\it The Brauer group of graded
continuous trace $C\sp *$-algebras}, Trans.\ Amer.\ Math.\
Soc., {\bf 308} (1988), no. 1, 115--132.

\bibitem[Qu]{Qu} D.~Quillen,
{\em Superconnections and the Chern character}, Topology
{\bf 24} (1985), no. 1, 89-95.

\bibitem[RW]{RW} I.~Raeburn and D.~Williams,
{\em Morita equivalence and continuous trace $C^*$-algebras},
Math.\ Surv.\ and Mono., {\bf 60} (1998).

\bibitem[Ros]{Ros} J.~Rosenberg, {\em Continuous trace algebras
from the bundle theoretic point of view}, Jour.\ Austr.\ Math.\ Soc.,
{\bf 47} (1989), 368--381.

\bibitem[SW]{SW} N.~Seiberg and E.~Witten, {\em String theory and
noncommutative geometry}, JHEP {\bf 09} (1999) 032, [{\tt hep-th/9908142}].

\bibitem[Se]{Se} A.~Sen, {\em Stable non-BPS states in string theory},
JEHP {\bf 06} (1998) 007, [{\tt hep-th/9803194}];
{\em ibid.}, {\em Stable non-BPS bound states of BPS $D$-branes},
JHEP {\bf 08} (1998) 010,\newline
[{\tt hep-th/9805019}]; {\em ibid.}, {\em Tachyon condensation on the brane
antibrane system}, JHEP {\bf 08} (1998) 012, [{\tt hep-th/9805170}];
{\em ibid.}, {\em $SO(32)$ spinors of type I and other solitons
on brane-antibrane pair}, JHEP {\bf 09} (1998) 023, [{\tt hep-th/9808141}].

\bibitem[Was]{Was} A.~Wassermann, {\em Automorphic actions of 
compact groups
on operator algebras}, Ph.D.\ dissertation, Univ.\ of Pennsylvannia, 
Philadelphia,
1981; {\em ibid.}, {\em Equivariant K-theory I: compact transformation groups
with one orbit type}, IHES preprint, 1985.

\bibitem[WO]{WO} N.E. ~Wegge-Olsen,
{\em $K$-theory and $C\sp *$-algebras:
A friendly approach}, (Oxford Science Publications,
The Clarendon Press, Oxford University Press, New York, 1993).

\bibitem[Wh]{Wh} G.~Whitehead,
{\em Elements of homotopy theory},
Graduate Texts in Mathematics {\bf 61}, (Springer-Verlag, New
York-Berlin, 1978).

\bibitem[Wi]{Wi} E.~Witten, {\em $D$-branes and $K$-theory},
JHEP {\bf 12} (1998) 019, [{\tt hep-th/9810188}].


\end{thebibliography}
\end{document}